\documentclass[12pt]{article}
\input epsf
\advance\vsize by 4.0 truecm
\begin{document}
\baselineskip16pt
\thispagestyle{empty}
\date{}
\title{Can Induced Gravity Isotropize \\
Bianchi I, V, or IX Universes?}
\author{J.L. Cervantes-Cota\thanks{e-mail:jorge@nuclear.inin.mx}\\
Departamento de F\'{\i}sica, \\
Instituto Nacional de Investigaciones Nucleares (ININ) \\
P.O. Box 18--1027, M\'exico D.F. 11801, M\'exico \\
Fax: +52-5-329 7301 \\[2pt]
and \\[2pt]
P. Ag. Chauvet\thanks{e-mail:pcha@xanum.uam.mx} \\
Departamento de F\'{\i}sica, \\
Universidad Aut\'onoma Metropolitana--Iztapalapa \\
P.O. Box 55--534, M\'exico D.F. 09340, M\'exico \\
Fax: +52-5-724 4611} 
\maketitle
\vskip 1 cm
%
%
\begin{abstract}
         
We analyze if Bianchi I, V, and IX models in the Induced Gravity (IG) theory 
can evolve to a Friedmann--Roberson--Walker (FRW) expansion due to 
the non--minimal coupling of gravity and the scalar field.  The analytical
results that we found for the Brans-Dicke (BD) theory are now applied 
to the IG theory which has $\omega \ll 1$ ($\omega$ being the square ratio 
of the Higgs to Planck mass) in a cosmological era in which the IG--potential 
is not significant.  We find that the isotropization mechanism 
crucially depends on the value of $\omega$.  Its smallness also permits 
inflationary solutions.    For the Bianch V model inflation due to the Higgs 
potential takes place afterwads, and subsequently the 
spontaneous symmetry breaking (SSB) ends with an effective FRW evolution.  
The ordinary tests of successful cosmology are well satisfied.
\end{abstract}

\newpage

\section{INTRODUCTION} 

One of the main problems of modern cosmology is to find a satisfactory 
explanation to both the small--scale inhomogeneity of matter distribution and 
the large--scale degree of isotropy measured in the Cosmic Microwave 
Background Radiation (CMBR) by the Cosmic Background Explorer (COBE) satellite 
\cite{Cobe94,KoTu90}.  Accordingly, one desires to construct a 
general model that explains both, antagonistic, properties of our Universe. In 
order to find such a solution, it is interesting to investigate if homogeneous, 
anisotropic (Bianchi) models can predict the level of isotropy detected by 
the COBE satellite and, at the same time, reproduce the local 
character of our Universe.  There have been various attempts to 
solve this problem (for instance, see Ref. \cite{Ba95}), but inflationary 
cosmologies are still the most appealing since they provide explanation to 
some other problems, as well.  Accordingly, if at its outset the Universe were
neither homogeneous nor isotropic, then because of a de Sitter stage, it will 
tend to homogeneity and isotropy.  Inflationary models, however, still
assume some fine tuned initial conditions and most of them 
have assumed a FRW symmetry from the outset: This is the first fine--tuning 
one invokes in doing cosmology, since from all possible set of initial 
conditions, a FRW Universe selects a very special set of homogeneous and 
isotropic space--time geometries.  Then, by consi\-de\-ring more general 
space--time symmetries the question arises, whether inhomogeneous and/or 
anisotropic cosmological models help to understand the naturalness of 
inflation.  In general relativity (GR) within anisotropic, Bianchi--type models 
it is claimed that a positive cosmological constant provides an effective 
means of isotropizing homogeneous Universes \cite{Wa83}.  The idea behind is 
that of the cosmic no--hair conjecture which states that in 
the presence of a cosmological constant the 
universe evolves into a de Sitter space-time \cite{nh-conj}, at least 
locally \cite{Sc93}.  The no--hair property ensures that all inhomogeneities 
will be smoothed out in a region of the event horizon.  The conjecture has 
been proved for a number of models \cite{nh-models},  where it was realized 
that is highly related to the homogenization and 
isotropization of cosmological models.  The initial conditions for inflation 
have been reviewed in Ref. \cite{GoPi92} and the situation is that 
inhomogeneous, anisotropic models with 
negative curvature fulfills the conjecture as well, but big 
initial inhomogeneities lead to the formation of black holes in some regions 
\cite{ShNaNaMa93}, however in other regions inflation is possible, achieving a 
physical scenario in which inflating regions are surrounded by black holes 
\cite {ChMo94}; this resembles the chaotic scenario of initial conditions 
\cite{Li83}. 

In general, it is suggested that a patch of the Universe should be at some 
level homogeneous to consider it as a right model where inflation can take 
place, otherwise inflation can be prevented.  Accordingly, before 
one regards inflation one should analyze the properties of the Universe, and 
see if some set of initial conditions will bring our Universe to a sufficient 
smooth patch to start the inflationary expansion.  This has motivated us to 
analyze if homogeneous, anisotropic models with a non-minimal coupling 
tend to loose its hairs.  Most of the results above apply to GR with perfect 
fluids minimally coupled (for non-perfect fluids see Ref. \cite{AnMaRoRy91}),  
and for non-minimally coupled fields see Refs. \cite{PiSc89,FuRoMa89,Capo95}, 
where the no--hair conjecture has been proved for some scalar tensor theories, 
including some particular potentials and various 
non-minimal couplings.  By trying to tacle this problem, we have shown 
that Bianchi type models (I, V, and IX) in the Brans-Dicke (BD) 
theory \cite{BrDi61} tend to isotropize as time goes on \cite{ChCe95}.  These 
models show an asymptotic FRW behavior, but only few with an inflationary 
stage.  However, inflation turns out to be an important, desirable 
feature to solve the above--mentioned problems.  Thus, in order to obtain 
an inflationary model with graceful exit, one usually 
introduces a potential term.  Therefore,  in the present 
investigation we consider an IG theory that includes a non-minimal 
coupling with gravity ({\it a la} BD) and a potential associated with the 
scalar field, which in our case is identified with a Higgs field.   Within 
this theory our cosmological scenario begins with an 
anisotropic expansion of Bianchi type I, V, or IX, but only type V 
evolution to a FRW model is consistent with imposed restrictions. 
In this way one achieves a sufficient smooth patch in the Universe 
preparing the `system' to be able to inflate, i.e., all physical fields present 
isotropize.   The isotropization mechanism occurs  in a similar way 
as in the BD theory \cite{ChCe95} where no potential exists.  However, in the 
present theory the isotropization process is inflationary, whereas in the BD 
theory it is not necessarily the case, see Ref. \cite{Ce98a}.  The effective  
equivalence of both theories is possible because, during the time of 
isotropization, the potential term in the IG theory will not contribute 
significantly as a stress energy to the 
dynamical equations, in our scenario\footnote{One can alternatively 
prepare the `system' in such a way that the potential term is at the very 
beginning the dominant contribution to the dynamical equations and, therefore, 
inflation occurs directly.  This is the ordinary scenario within which 
inflation takes automatically place and the no--hair conjecture is 
fulfilled.\label{f2}}.  Afterwards, because of the Higgs potential of the 
theory, during the SSB of the Higgs field, a second inflationary 
era follows.  Finally, after inflation a FRW behavior is dominant.

This work is organized as follows: In section \ref{igbdgr} we present the IG 
theory, pointing out some differences among it, BD, and GR theories.  In 
section \ref{anibh} the cosmological field equations are analyzed in view of 
the isotropization of the solutions, employing some results for the BD theory 
that turn out to be valid for this theory, as well.  In section \ref{physce} 
we present the above-mentioned physical scenario.  Finally, the conclusions 
are written in section \ref{coniso} .

\section{IG, BD, AND GR THEORIES \label{igbdgr}}

We have investigated two viable IG models of inflation, one using the SU(5) 
Higgs field \cite{CeDe95a} and, the other, the SU(2) Higgs \cite{CeDe95b}, 
both coupled non--minimally to gravity.  The Lagrangian for both
theories has the same mathematical form, and therefore, the qua\-li\-tative 
behavior of the cosmological models are very similar.   In the present work, we 
consider the specific physical scenario using the SU(5) Higgs field--gravity 
theory, fully discussed in Refs. \cite{CeDe95a,DeCo}, however most of the 
conclusions are valid also for the SU(2) Higgs field--gravity model.  The 
Lagrangian of the IG theory is, with signature (+,--,--,--),
\begin{equation}
\label{lag}
{\cal L}= \left[ \frac{1}{8 \omega} 
\mbox{$ {\rm tr} \Phi^{\dag} \Phi$} \, R + \frac{1}{2}
 {\rm tr} D_{\mu}\Phi^{\dag} D^{\mu} \Phi 
- V(\mbox{$ {\rm tr} \Phi^{\dag} \Phi$}) +  L_{M} \right] \sqrt{- g} \, ,
\end{equation}
where greek indices denote space-time components, $R$ is the Ricci scalar, 
and $\Phi$ is the $SU(5)$ isotensorial
Higgs field.  The symbol $D_{\mu}$ means the 
covariant gauge derivative with respect to all gauged groups: 
$D_{\mu} \Phi = \Phi_ {| \mu } + i {\rm g_{5}} [A_ \mu , \Phi]$, where
$A_{\mu} = A_ \mu {}^a \tau _a $ are
the gauge fields of the inner symmetry group, $\tau_a$  are its generators,
and g$_{5}$ is the coupling constant of the gauge group
($|\mu $ means the usual partial derivative). $\omega$ is a   
dimensionless, coupling constant parameter that regulates the strength of 
gravitation and $ L_{M}$ contains only the fermionic and massless bosonic 
fields, which belong to the inner gauge group $SU(5)$; and the Higgs potential 
is given by
\begin{equation}
\label{eqph}
V(\mbox{$ {\rm tr} \Phi^{\dag} \Phi$}) = \frac{\mu ^2}{2}
\mbox{$ {\rm tr} \Phi^{\dag} \Phi$} + \frac{\lambda }{4!}
(\mbox{$ {\rm tr} \Phi^{\dag} \Phi$})^2 +
\frac{3}{2} \frac{\mu ^4}{\lambda } =  
\frac{\lambda}{24} \left(\mbox{$ {\rm tr} \Phi^{\dag} \Phi$}  
+ 6 \frac{\mu^2}{\lambda} \right)^2 \,\ ,
\end{equation}
where we added a constant term to prevent a negative cosmological
constant after the SSB \footnote{From the particle physics point of view, it
is not suggested to add a cosmological constant, but is neither forbidden
\cite{We89}; this constant is 
$\Lambda = 12 \pi G \frac{\mu ^4}{\lambda} \sim 10^{21}$ GeV$^{2}$.  However, 
in cosmology fitting the dynamics of cluster of galaxies suggests (in GR) that 
$\Omega_{\Lambda} \sim {\cal O} (1)$ \cite{KoTu90}, that is, 
$\Lambda \sim 10^{-83}$ GeV$^{2}$; this is the heart of the 
cosmological constant problem.}.  Because of the presence of mass terms in
Eq. (\ref{eqph}), the Lagrangian Eq. (\ref{lag}) is not conformally  
invariant with GR, cf. \cite{BiDa82}; this is important to mention 
because there are a
number of results using conformal transformations among different theories 
demanding the same physics, but in our case such transformations are not 
conformally invariant.

The field $\frac{2\pi}{\omega} {\rm tr} \Phi^{\dag} \Phi$ plays the role of the 
inverse of Newton's gravitational constant ($G^{-1}$) and after a SSB process, 
when the $\Phi-$field becomes a constant,
${\rm tr} \Phi^{\dag} \Phi=-6\frac{\mu^{2}}{\lambda}$, the potential vanishes.  
In this way, after the SSB this theory becomes {\it effectively} GR.  Further, 
some fermions and boson fields that become 
massive after the breaking appear as a source's contribution to the right 
hand side of Einstein equations, for details see Refs. \cite{CeDe95b,DeCo}.

From Eq. (\ref{lag}), the gravity field equations are:
\begin{eqnarray}
\label{eqrp}
R_{\mu \nu} -\frac{1}{2} R g_{\mu \nu}  \, & = & \,
- {4 \omega \over \mbox{$ {\rm tr} \Phi^{\dagger} \Phi $}}   
\left[ T_{\mu \nu} + V(\mbox{$ {\rm tr} \Phi^{\dagger} \Phi $}) \, 
g_{\mu \nu} \right] \nonumber\\ &&
 - {4 \omega \over \mbox{$ {\rm tr} \Phi^{\dagger} \Phi $}} 
 \left[{\rm tr} D_{(\mu} \Phi^{\dagger} D_{\nu )} \Phi   -
\frac{1}{2} \, {\rm tr} D_{\lambda} \Phi^{\dagger} D^{\lambda} 
\Phi \,\  g_{\mu \nu} \right]
 \nonumber\\ &&
- \frac{1}{\mbox{$ {\rm tr} \Phi^{\dagger} \Phi $}} 
\left[ (\mbox{$ {\rm tr} \Phi^{\dagger} \Phi $})_{| \mu || \nu} -
(\mbox{$ {\rm tr} \Phi^{\dagger} \Phi $})^{| \lambda}_{ \,\  \,\ || \lambda} 
\,\ g_{\mu \nu} \right] 
\,\ , 
\end{eqnarray}
where $ T_{\mu \nu} $ is the energy--momentum tensor belonging to 
$L_M \sqrt{-g}$ in (\ref{lag}) alone, $|| \mu$ is the usual covariant 
derivative, and the Higgs field equation is 
\begin{equation}
\label{eqh}
\left( D^{\lambda} \Phi \right)_{|| \lambda} + 
\frac{\delta V}{\delta \Phi^{\dagger} } - \frac{1}{4 \omega} R \Phi \,\ = \,\
2 \frac{\delta L_{M} } {\delta \Phi^{\dag}}  \,\ = \,\ 0 \,\ , 
\end{equation}
where the contraction of the double covariant derivative is 
$\left( D^{\lambda} \Phi \right)_{|| \lambda} = D_{\lambda} D^{\lambda} \Phi + 
\Gamma^{\lambda}_{\,\ \mu \lambda} D^{\mu} \Phi$.  If there were any Yukawa 
couplings, this equation would not be equal to zero (this is actually the case 
for the Standard Model of Particle Physics, see Ref. \cite{CeDe95b}).

Mathematically, IG and BD theories are equal except for the potential and  
the meaning of covariant derivatives.  This can be seen by identifying 
$\phi=\frac{2 \pi}{\omega} \mbox{${\rm tr} \Phi^{\dagger} \Phi$}$, where 
$\phi$ is the BD field.  Indeed, BD and IG theories are related.  The idea 
to induce gravity by a Higgs field has been already discussed elsewhere 
\cite{ig}, and the motivation for us is that
the field coupled to the matter content of the Universe, 
{\it a la} Brans and Dicke \cite{BrDi61}, is the same that produces   
their masses, i.e., a Higgs field.  Then, the identification of both
scalar fields is very appealing, see Refs. \cite{DeFrGh,CeDe95a}.  Though this 
identification is quite simple, the resulting IG theory presented above is 
more elaborated than the BD theory.  Accordingly, in the IG theory the 
$\Phi-$field is a Higgs field with its 
associated potential. Then, a matter content appears explicitly with its 
corresponding energy scales.  In fact, there are three energy scales
to deal with: the Planck, the Higgs, and the $X$ boson masses.  The Higgs mass 
energy is given through Eq. (\ref{eqh}) [see also Eq. (\ref{hig}) below], 
$M_{H}=- \left( \frac{4 \omega}{3+2 \omega} \right) \mu^{2}$, and it 
determines the dynamical behavior of the $\Phi-$field once the 
SSB begins to occur.  A second 
energy scale is given by the $X$ (the same as the $Y$) boson mass, 
$M_{X}=\sqrt{10 \pi} g_{5}\frac{\mu}{\sqrt{\lambda}}\approx 10^{15}$ GeV.  
Finally, the Planck energy scale is given through $M_{Pl}\equiv \sqrt{2G}$.  
After the SSB 
$\frac{2\pi}{\omega} \mbox{${\rm tr} \Phi^{\dagger} \Phi$} =  
\frac{1}{G}$ implying that the coupling constant must be 
$\omega = - \frac{6\pi}{\lambda}\left(\frac{\mu}{M_{Pl}}\right)^{2}
\approx 10^{-6}$, which is very small because one is forcing to match two 
energies scales given by the Planck and boson masses through the non-minimal 
coupling in Eq. (\ref{lag}).  

In contrast to what happens in the IG theory, in 
the BD theory there is no potential nor is GR induced after a SSB process.  
Therefore, the $\phi-$field in BD should be nowadays a cosmic, scalar function, 
and to fit well the theory with the experimental data the coupling constant 
must have a great value \cite{Wi93}, $\omega>500$, making BD and GR theories 
very similar.

In section \ref{physce}, we will present a cosmological scenario in the 
IG theory, yet employing some results found for the BD theory which turn out to 
be also
valid for the IG theory.  As pointed out above, both theories have the same 
mathematical form when the potential term plays no significant role in the 
field equations.  Therefore, in order to translate analytic BD results to the 
IG theory and to clarify when this situation is correct, we put the 
above equations in terms of the BD field 
$\phi=\frac{2 \pi}{\omega} \mbox{${\rm tr} \Phi^{\dagger} \Phi$}$, which in 
our case represents the excited Higgs field.  Then, the IG gravity equations 
are now,
\begin{eqnarray}
\label{gr}
R_{\mu \nu} -\frac{1}{2} R g_{\mu \nu}  \, & = & \,
-  \frac{8 \pi}{\phi} \left[ \hat{T}_{\mu \nu} + V(\phi) \, g_{\mu \nu} \right]
\nonumber\\ &&
 - \frac{\omega}{\phi^{2}}  
 \left[ \phi_{| \mu} \phi_{| \nu}   -
 \frac{1}{2} \phi_{| \lambda} \phi^{| \lambda} \,\ g_{\mu \nu} \right]
 \nonumber\\ &&
- \frac{1}{\phi} 
\left[ \phi_{| \mu || \nu} -
\phi^{| \lambda}_{ \,\  \,\ || \lambda} \,\ g_{\mu \nu} \right] 
\,\ , 
\end{eqnarray}
and the Higgs field equation is  
\begin{equation}
\label{hig}
 \phi^{| \lambda }{}_{|| \lambda} - \frac{4 \omega}{3+2\omega} \, 
 (\phi - G^{-1}) \,\ = \,\ \frac{8 \pi}{3+2\omega} \, \hat{T} \,\ ,
\end{equation}
where $\hat{T}$ is the trace of the {\it effective} energy--momentum tensor, 
$\, \hat{T}_{\mu \nu }$, given by
\begin{equation}
\label{eqtt}
\, \hat{T}_{\mu \nu} \,\ = \,\ T_{\mu \nu} +
\frac{G}{4 \pi} \, \phi \,\ M_{ab}^{2}
\left( A^{a}_{\,\ \mu} A^{b}_{\nu} - \frac{1}{2} g_{\mu \nu}
A^{a}_{\,\ \lambda} A^{b \lambda}   \right) ,
\end{equation}
where $M^{2}_{ab}$ is the gauge boson mass square matrix, stemming from the 
covariant gauge derivatives [see discussion after Eq. (\ref{lag})].

The continuity equation (energy--momentum conservation law) reads
\begin{equation}
\label{eqct}
\, \, \hat T^{\,\ \nu}_{\mu \,\ \,\ || \nu} \,\ = \,\ 0 ,
\end{equation}
and in the present particle physics theory, $SU(5)$ GUT, 
all the fermions remain massless after the first symmetry--breaking and no 
baryonic matter is originated in this way.  This is the reason to have 
Eq. (\ref{eqh}) equal to zero, too.

The source term in Eq. (\ref{hig}) is important for reheating, since the
Higgs field remains coupled to $\hat{T}$, i.e., to gauge boson fields.  In
Ref. \cite{SaBoBa89} is claimed that there are no couplings between the Higgs
field and other fermionic or bosonic fields, but in our induced gravity
approach there indeed exist bosonic field's couplings.

Eqs. (\ref{gr},\ref{hig},\ref{eqtt},\ref{eqct}) are the field equations for 
the IG theory written
in terms of the BD field.  These equations would describe the BD theory if the 
potential in Eq. (\ref{gr}) vanishes, implying that the second term
on the left hand side of Eq. (\ref{hig}) vanishes, and if the second term on 
the right hand side of Eq. (\ref{eqtt}) vanishes, as well. Then, by bringing 
BD analytic results to the IG theory, one has to be sure that these 
conditions apply.  

The above equations reduces to the GR equations once the SSB takes 
place, when the $\phi-$field becomes a constant and the potential vanishes.

Next, we consider anisotropic universes and study their asymptotic behavior.

\section{ANISOTROPIC MODELS AND ASYMPTOTIC BEHAVIOR \label{anibh}}

We consider homogeneous, aniso\-tro\-pic Bianchi type models that could 
experience, at least in principle, an isotropization mechanism evolving 
to a FRW model.  Therefore, we study the dynamics of Bianchi type I, V, and 
IX spacetime symmetries in a synchronous coordinate frame; a general 
discussion of Bianchi models is found in Ref. \cite{Ma79}.

In order to translate the results of Ref. \cite{ChCe95} to IG, we will put 
the cosmological field equations in terms of the following scaled variables 
and definitions: the scaled Higgs field $\psi \equiv \phi a^{3(1-\nu)}$, a new 
cosmic time parameter 
$d\eta \equiv a^{-3\nu} dt$, $()^\prime\equiv \frac{d}{d\eta}$, 
the `volume' $a^{3}\equiv a_{1}a_{2}a_{3}$, and the Hubble parameters 
$H_{i}\equiv {a_{i}}^\prime /a_{i}$ corresponding to the scale factors 
$a_{i}=a_{i}(\eta)$ for $i=1,2,3$.  One can assume a barotropic equation 
of state for the perfect fluid represented by $\hat{T}_{\mu \nu}$, 
$p=\nu \rho$, with $\nu$ a constant. Using these definitions 
and the above--mentioned metrics, one obtains the cosmological equations 
from Eqs. (\ref{gr}--\ref{eqct}):
\begin{equation}
\label{a123}
(\psi H_i)^\prime -  \psi a^{6 \nu} C_{ij}  \, = \, 
\frac{8 \pi a^{3(1+\nu)}}{3+2\omega}  \left[[1 + (1 - \nu) \omega] \rho + 
(3+2\omega) V + \frac{\delta V }{G} \right] 
 ~~~ {\rm for} ~~~ i=1,2,3. 
\end{equation}

\begin{eqnarray}
\label{h123} 
&&H_{1}H_{2} + H_{1}H_{3} + H_{2}H_{3} + 
[1+(1-\nu)\omega] \, \left(H_{1}+H_{2}+H_{3}\right) 
\frac{\psi^\prime}{\psi}
\nonumber \\[2pt]
&&- (1-\nu)[1+\omega(1-\nu)/2] (H_{1}+H_{2}+H_{3})^{2} 
- \frac{\omega}{2} \left( \frac{\psi^\prime}{\psi}\right)^{2} 
- \frac{C_{j}}{2} a^{6 \nu}  \nonumber \\[2pt]
&& = \,\ 8 \pi \, \frac{a^{3(1+\nu)}}{\psi} \, [\rho + V] \,\ ,
\end{eqnarray}

\begin{eqnarray}
\label{psi} 
&&\psi^{\prime \prime}  + (\nu-1) a^{6 \nu} C_{j} \, \psi =  \nonumber \\[3pt]
&&\frac{8 \pi a^{3(1+\nu)}}{3+2\omega} \left[ [2(2-3\nu)+3(1-\nu)^{2} \omega ] 
\rho + 3(1-\nu)(3+2\omega)V + (1-3\nu) \frac{\delta V}{G}   \right] \,\ ,
\end{eqnarray}
and
\begin{equation}
\label{eqctsol}
\rho a^{3(1+\nu)} = {\rm const.} \equiv M_{\nu} \,\ ,
\end{equation}
where $\delta V \equiv 
\frac{\partial V}{\partial \psi} \frac{\partial \psi}{\partial \phi}$ and 
$C_{j}\equiv \Sigma_{i} C_{ij}$ is the curvature corresponding to
different $j$--Bianchi models (j=I, V, or IX).  The subscript $i=1,2,3$ refers
to the three scale factors. Accordingly, one has that
\begin{equation}
\label{curv}
\matrix {
& &~{}^I &~~~~~~~~~{}^V~~~~~~~~ &~{}^{IX} \cr 
& & & & \cr
& &0 & \frac{2}{a_1^2} & \frac{a_1^4-a_2^4-a_3^4 + 2 a_2^2 a_3^2}{- 2 a^6} \cr
& & & & \cr
& C_{ij} \,  \equiv &0 &  \frac{2}{a_1^2} & 
\frac{a_2^4-a_3^4-a_1^4 + 2 a_1^2 a_3^2}{- 2 a^6}  \cr 
& & & & \cr
& &0 & \frac{2}{a_1^2} & \frac{a_3^4-a_1^4-a_2^4 + 2 a_1^2 a_2^2}{- 2 a^6} 
\, . \cr} 
\end{equation}

Equations (\ref{a123}, \ref{h123}, \ref{psi}, \ref{eqctsol}) form the complete 
set of equations to be integrated.  For the Bianchi V model there is 
additionally
the following constriction
\begin{equation}
\label{h12}
H_{2} + H_{3} \,\ = \,\ 2 H_{1}   \,\ ,
\end{equation}
implying  that $a_{2}$ and $a_{3}$ are inverse proportional functions, 
$a_{2}a_{3}= a_{1}^{2}$.

We study in the following only the anisotropic character of the  
solutions, and not the influence of the potential of the theory.  Otherwise, 
the potential term will automatically produce an inflationary stage from its 
outset (see footnote \ref{f2}; cf. Ref. \cite{Wa83}), and what we desire 
is to have a model in which inflation takes place only after the 
isotropization process has almost concluded, up to 
some extent at least.  This would guarantee 
that anisotropic stresses decrease with  time, as it is the case in GR 
\cite{Wa83,nh-models}.   We want to investigate the dynamics 
before inflation occurs to see if the 
model dynamically tends to a FRW model.  If this were the case, any physical 
perturbation (hairs) present will experience an isotropization mechanism 
resulting in the smoothing of any patch of the Universe.  For instance, let us
assume there exist additionally other fields (dilaton, matter fields, etc.), 
whose stress energies do not contribute significantly to the dynamical 
processes, at least for some time interval\footnote{Remind that any field, 
governed by its field equation, has an inherent typical time determined by its 
mass scale, or by some constant of nature involved in its 
field equation.  Normally, if there are many 
fields present one expects every field to be significant for the evolution in 
some characteristic time scale.}.  Then, the Universe dynamics governed by the 
$\phi-$field will isotropize all these extra fields.  Thus, thinking in a 
chaotic scenario where the initial conditions for inflation imply that the 
region, and the inflaton field itself, should be sufficiently homogeneous and 
isotropic, then, after such an isotropization process it is more likely that 
inflation takes place successfully.  Therefore, an isotropization mechanism can 
be important in pre-inflationary dynamics.

Inflation is a nice feature to solve the problems of Standard Model of 
cosmology, but most realistic models of inflation \cite{KoTu90,Ol90,Li90} 
are fine tuned.  For instance, a fact that is usually omitted is 
that to achieve enough $e$--foldings of expansion new inflationary models 
demand the initial inflaton field (say, $\varphi$) to have very small values, 
about $\varphi_{o} < 10^{-5} v$, where $v$ is the true vacuum value of the 
$\varphi-$field.  This fact can be understood with the help of the slow 
rollover conditions: $-V''<9 H^{2}$ and 
$\left(\frac{V'}{V} \right)^{2}<48 \pi G$, which in turn imply, respectively, 
that $v \, {}^{>}_{\sim} \, M_{Pl}$ and $\varphi/v < v / M_{Pl}$.  Typically, 
GUT theories have $v \sim 10^{14\--15}$GeV, then the second condition
implies very small initial values for $\varphi$, whereas the first condition
is a severe impediment (or inconsistency with realistic particle physics) to 
have enough $e$--foldings of expansion, and hence, to solve
the horizon and flatness pro\-blems of cosmology.  Another fine--tunning aspect
or, to say precisely, inconsistency relies on the fact that $\lambda<10^{-12}$ 
[$\lambda$ coming from a potential similar to Eq. (\ref{eqph})] to fit well the 
temperature fluctuations measured by the COBE satellite \cite{Cobe94}.  Yet 
from particle physics one expects that $\lambda \sim 1$ (in any case not 
that small as required above!).  Further, such smallness of $\lambda$ works in 
opposite sense as 
for producing a high reheating temperature ($T_{RH}$) after inflation, since 
typically $T_{RH} \sim \lambda^{1/4} \, v$.  Then, the baryon asymmetry could 
not be attained, unless very fine tuning initial conditions are chosen.  
Therefore, it is interesting to investigate models that achieve a successful 
inflationary stage when they do not start with standard, inflationary initial 
conditions.   Accordingly, our motivation is to study cosmological scenarios 
in scalar tensor theories with a particle physics content, and to consider 
more general initial conditions to understand some of the ac doc assumptions or
problems of inflationary cosmologies.  An important issue is naturally the 
study of cosmological isotropization processes.

We return to our model in which the isotropization mechanism occurs before 
inflation.  Accordingly, one must guarantee that potential terms in Eqs. 
(\ref{a123}, \ref{h123}, \ref{psi}) are less signi\-fi\-cant than the perfect  
fluid term (given through $\rho$).  These conditions imply respectively that:
\begin{eqnarray}
\label{conds}
&& [1+(1-\nu)\omega] \rho > \frac{3+2\omega}{16\pi} M_{H}^{2} M_{Pl}^{2} \, 
[(3+2\omega)(\phi G-1)^{2} + 2 (\phi G-1)] \ ,\,
\nonumber \\[2pt]
&&\rho>V(\phi) = \frac{3+2\omega}{16\pi} M_{H}^{2} M_{Pl}^{2} \, (\phi G-1)^{2} 
\, , \, {\rm and} \nonumber \\[2pt]
&&(1-3\nu)\rho>-\frac{3+2\omega}{4\pi} M_{H}^{2} M_{Pl}^{2} \, (\phi G-1) \,\ .
\end{eqnarray}
The first condition is the most 
restrictive, but it suffices to have  
$\rho \, {}^{>}_{\sim} M_{H}^{2} M_{Pl}^{2}$ for\footnote{Note that $\phi G>1$ 
is equivalent to 
$\mbox{$ {\rm tr} \Phi^{\dag} \Phi$} > -6 \frac{\mu^2}{\lambda}$. \label{f5}} 
$\phi G>1$, which is not a severe condition at all.  Under these assumptions 
the IG cosmological equations are {\it effectively} equivalent to 
the BD cosmological equations.  Therefore, we are able to employ the analytic 
solutions found for the Bianchi I, V, and IX models in the BD theory 
\cite{ChCe95} on the IG theory.   These solutions are valid during the time 
interval the above inequalities apply, say, from the initial 
time $\eta_{o}$ to $\eta_{1}$.

In order to analyze the anisotropic character of the solutions, we have 
constructed the following `constraint' equation \cite{ChCe95} using the BD 
equations analogous to Eqs. (\ref{a123}, \ref{h123}, \ref{psi}), valid from the
time $\eta_{o}$ to $\eta_{1}$:
\begin{eqnarray}
&& \sigma ( \eta) \equiv ~ - (H_1 - H_2)^2 - (H_2 - H_3)^2 - (H_3 - H_1)^2  
\,\ = \,\ \nonumber \\[4pt]
&&{3 \over {2(1-\nu)}}\left( {\psi^{\prime \prime} \over \psi} \right)-
{1\over{(1-\nu)^2}}\left( {\psi^{\prime} \over \psi} \right )^2 -
{{(1-3 \nu)}\over{(1-\nu)^2}} \left( {{ (1-3 \nu)m_\nu\eta + \eta_o}
\over \psi} \right) \left( {\psi^{\prime} \over \psi} \right )\nonumber \\[4pt]
&& +{{[2-3 \nu+ {3\over2}\omega(1-\nu)^2]}
\over {(1-\nu)^2}} \left({{ (1-3 \nu)m_\nu\eta + \eta_o}
\over \psi}\right)^2+{{3 [2+\omega(1-\nu)(1+3\nu)] m_\nu}
\over{2(1-\nu) \psi}} 
\, . \label{con2}  \end{eqnarray}
$\sigma$ is the anisotropic shear. $\sigma = 0$ is a
necessary condition to obtain a FRW cosmology since it implies $H_1 = H_2 =
H_3$, cf. Ref. \cite{ChCeNu91}. If the sum of the squared
differences of the Hubble expansion rates tends to zero, it would mean that the 
anisotropic scale factors tend to a single function of time which is,
certainly, the scale factor of the FRW models.

We have shown elsewhere \cite{ChCe95} that 
$ \psi \,= \,A_{j} \eta^2 + B_{j} \eta + C_{j} $ is a solution for the 
homogeneous, anisotropic models, where $A_{j}$, $B_{j}$, and $C_{j}$ are 
some constants depending on the $j$--Bianchi type.  The Hubble expansion 
rates are given through 
\begin{eqnarray}
\label{hi}
&&H_{1}+H_{2}+H_{3} \, =  \, 
\frac{1}{(1-\nu)} \frac{[2 A_{j} - \frac{8 \pi M_{\nu}}{3+2\omega}(1-3\nu)] 
\eta + B_{j} - \eta_{o}}{\psi}  \, ,  \nonumber  \\[4pt]
&&H_i \, = \, {1 \over 3} \left( H_{1}+H_{2}+H_{3} \right) + {h_i \over \psi} 
\, ,
\end{eqnarray}
where the $h_i$'s are functions that 
determine the anisotropic character of the solutions and are intimately 
related to the constants $A_{j}$, $B_{j}$, and $C_{j}$ as follows
\cite{RuFi75,ChCe95}:

\bigskip

{\bf Bianchi type I}:
\begin{eqnarray}
\label{b1ctes}
& A_{{}_{I}} = &  [2-3 \nu + \frac{3}{2} \, \omega \, (1-\nu)^{2}] 
 \, m_{\nu}  \nonumber \\
& C_{{}_{I}} = & \frac{- 3 (1-\nu)^{2} \left( h_{1}^{2}+h_{2}^{2}+ h_{3}^{2} 
\right)/2 
+ (1-3\nu)  \eta_{o} B_{{}_{I}} +B_{{}_{I}}^{2} 
- \left( 2-3 \nu + \frac{3}{2}\omega(1-\nu)^{2} \right) \eta_{o}^{2}}
{3 m_{\nu}(1-\nu)^{2} (3+2\omega)} 
\end {eqnarray}
where the $h_{i}$ are constants and $B_{{}_{I}}$ remians a free parameter.  

\bigskip

{\bf Bianchi type V}:
\begin{eqnarray}
\label{b5ctes}
& A_{{}_{V}} =& - \frac{(1-3\nu)^{2}}{1+3\nu} \, m_{\nu} \nonumber \\
& B_{{}_{V}} =& - 2 \, \frac{1-3\nu}{1+3\nu} \, \eta_{o} \nonumber \\
& C_{{}_{V}} =& \frac{-1}{m_{\nu}(1+3\nu)}   \left[ \frac{(1+3\nu)^{2} 
(h_{1}^{2}+h_{2}^{2}+h_{3}^{2})}{18 \nu + \omega (1+3\nu)^{2}} + \eta_{o}^{2} 
\right]  \nonumber
\end {eqnarray}
where  $h_{1}=0$ in accordance with Eq. (\ref{h12}), and $h_{2}$ ($=-h_{3})$ is a constant.

\bigskip

{\bf Bianchi type IX}:
 
In this case the $h_{i}$ are functions, $h_{i}=h_{i}(\eta)$, obeying the equation:
\begin{equation}
\label{hi5}
 h_i^{\prime}~=~a^{6 \beta} \psi C_{{}_{i\,IX}}~+~{{2A_{{}_{IX}} 
-[2(2-3 \beta)+3(1- \beta)^2 \omega] m_{{}_ \beta}} \over{3(1- \beta)}}
~~, \quad {\rm i=1,2,3} \end{equation}
subject to the condition
\begin{equation}
\label{hiquad}
h^2_1+h^2_2+h^2_3 \equiv~{ \rm K^2} = -{\omega^3\over 2(1-\beta)^2} 
\left[ P\eta^2+Q\eta+S\right] ~~~,  
\end{equation}
where the constants $P$, $Q$, and $S$, given in terms of $A_{{}_{IX}}$, 
$B_{{}_{IX}}$ and $C_{{}_{IX}}$, stand for
\begin{eqnarray}
\label{pqs}
&P =& XA_{{}_{IX}} - [4A_{{}_{IX}} - Y](1-3\beta)^2m_\beta~~~, \nonumber \\
&Q =& XB_{{}_{IX}} - [4A_{{}_{IX}}\eta_0-2Y m_\beta\eta_0 + 
2(1-3\beta)B_{{}_{IX}}](1-3\beta)m_\beta~~~,  \nonumber \\
&S =& XC_{{}_{IX}} - [2\Delta+ 2(1-3\beta)m_\beta \eta_0 B_{{}_{IX}} 
- Y m^2_\beta \eta^2_0]~~~, \nonumber \\
&X \equiv& 3(1+3\beta)(1-\beta)^2\omega m_\beta + 6(1-\beta)m_\beta-
2(1+3\beta)A_{{}_{IX}} ~~, \nonumber \\
&Y \equiv& 2(2-3\beta)+3(1-\beta)^2\omega~~~.  
\end{eqnarray}

The $h_{{}_i}$'s can be further given as
\begin{equation}
\label{h19}
h_1 =-\left[{\kappa^2 +4\kappa +1\over3( \kappa^2 + \kappa + 1)}
\right]{\rm K} ~~~, \end{equation}

\begin{equation}
\label{h29}
h_2 = \left[{-\kappa^2 +2\kappa +2\over3( \kappa^2 + \kappa + 1)} \right]
{\rm K} ~~~, \end{equation}
and
\begin{equation}
\label{h39}
h_3 = \left[{2\kappa^2 +2\kappa -1\over3( \kappa^2 + \kappa + 1)} \right] 
{\rm K} ~~~,
\end{equation}
where $\kappa$ is an unknown function of $\eta$.  
Unfortunately, we have not achieved yet to obtain the explicit functional 
dependence of $ \kappa = \kappa(\eta)$.  The axisymmetric 
case ($a_{1} = a_{2} \neq a_{3}$),  
assuming a quadratic function for $\psi$, gives the closed FRW solution,  
implying that $B_{{}_{IX}} = C_{{}_{IX}} = 0$.

\bigskip

The above--presented Bianchi models obey the condition: 
\begin{equation}
\label{resth123}
 h_1 + h_2 + h_3 = 0 \, , 
\end{equation}
then, the shear, Eq. (\ref{con2}), becomes
\begin{equation}
\label{sig}
\sigma(\eta) = - \frac{3 (h_{1}^{2} + h_{2}^{2} + h_{3}^{2})}{\psi^2}
\,\  .
\end{equation}
This equation admits solutions such that  $\sigma \to 0$ as $\eta \to \infty$  
(or $t \to \infty$), that is, one has time asymptotic isotropization 
solutions, similar to the solutions found for Bianchi models in GR, 
see Ref. \cite{BaSo86}.  In fact, one does not need to impose an 
asymptotic, infinity condition, but just that $\eta \gg \eta_{*}$, 
where $\eta_{*}$ is yet some arbitrary value, to warrant that $\sigma$ can 
be bounded from above.   For the Bianchi type IX 
$h_{1}^{2} + h_{2}^{2} + h_{3}^{2}$, given by Eq. (\ref{hiquad}), is not a 
constant but a quadratic function of $\eta$, however, the denominator 
of  Eq. (\ref{sig}) is a quartic polynomial in $\eta$, therefore, an 
asymptotic isotropic behavior, similar to the other models, is also expected.

The analytic flat, open and closed FRW solutions are obtained if $h_i=0$, for 
the Bianchi type I, V and IX, respectively.  In this case, it implies that 
$B_{j}=C_{j}=0$ for all the Bianchi models considered here.

\section{A PHYSICAL SCENARIO \label{physce} }

We consider in the following a scenario in which the   
isotropization process can occur from the very beginning, $\eta_{o}$, until 
the time $\eta_{*}$, corresponding to a time scale before inflation happens.  
That is,  one has that 
$\eta_{*} \leq \eta_{1}$, where $\eta_{1}$ is the time when inflation starts  
because the potential stress energy begins to be the major contribution
to Eqs. (\ref{a123}, \ref{h123}, \ref{psi}).  In this way, the 
isotropization of hairs is guaranteed indeed before the de Sitter 
stage occurs.

The integration of Eq. (\ref{hi}) to get explicitly the scale factor functions 
is straightforward, and was reported in Refs. \cite{RuFi75,ChCe95}.  The 
solutions are characterized by the sign of the discriminant, 
$\Delta_{j} \equiv B_{j}^{2} - 4 A_{j} C_{j}$, which implies two different 
behaviors depending on it being positive or not.  The 
solutions $\Delta > 0$ are restricted to be valid in a specific time 
interval\footnote{The solutions with $\Delta>0$ are qualitatively 
$a_{i} \sim e^{\rm arctanh\, \eta}$ valid for $-1 < \eta < 1$, whereas the 
solutions with $\Delta < 0$ are $a_{i} \sim e^{\rm arctan\, \eta}$ valid for 
$- \infty < \eta< + \infty$. \label{del}} and, additionally, they are 
asimptotically anisotropic. The solutions $\Delta \le 0$ are valid during 
the whole  cosmic time ($\eta$) interval.  Independent of the initial 
value the Hubble parameters may have (including $H_i < 0$), because of 
Eq. (\ref{sig}), as $\eta \rightarrow \eta_{*}$,  
$\sigma \rightarrow \sigma_{\rm min} \approx 0$, where the value of $\eta_{*}$ 
is fixed by the degree of isotropy ($\sigma_{\rm min}$) at that time in each 
Bianchi model.   

We analyze the conditions for this scenario to be viable on 
the Bianchi I, V, and IX models.
\bigskip

{\bf Bianchi I}

The discriminant of this model is given by \cite{RuFi75}:
\begin{eqnarray}
\label{discri1}
&\Delta_{{}_{I}} =& B_{{}_{I}}^{2} - 4 A_{{}_{I}} C_{{}_{I}} = 
B_{{}_{I}}^{2}-\frac{4[2-3\nu+\frac{3}{2}\omega (1-\nu)^2]}{3(1-\nu)^{2} 
(3+2 \omega)} \times \\  
&&\left[B_{{}_{I}}^{2} + (1-3\nu) \eta_{o} B_{{}_{I}} -
\frac{3}{2} (1-\nu)^{2} (h_{1}^{2}+h_{2}^{2}+h_{3}^{2}) -
[2-3\nu+\frac{3}{2}\omega (1-\nu)^2]\eta_{o}^{2}   \right] \, .\nonumber 
\end{eqnarray}
In our case, $\omega \ll 1$, and $\Delta_{{}_{I}}$ is always positive.  Therefore, 
here the isotropization is not possible, and the physical scenario fails. 

\bigskip

{\bf Bianchi V}

The discriminant of this model is given by \cite{ChCe95}:
\begin{equation}
\label{dis5}
\Delta_{{}_{V}} ={B_{{}_V}^2-4A_{{}_V}C_{{}_V}} = {{-8 (1-3\nu)^2} \over 
{18 \nu+(1+3\nu)^2 \omega}}{h_2^2}  \,\ . 
\end{equation}  
which implies $\Delta_{{}_{V}} < 0$.   In this case, the solutions isotropize 
and the scenario is successfully achieved. 

\begin{figure}
\unitlength 1in
\epsfxsize=5in \epsfbox{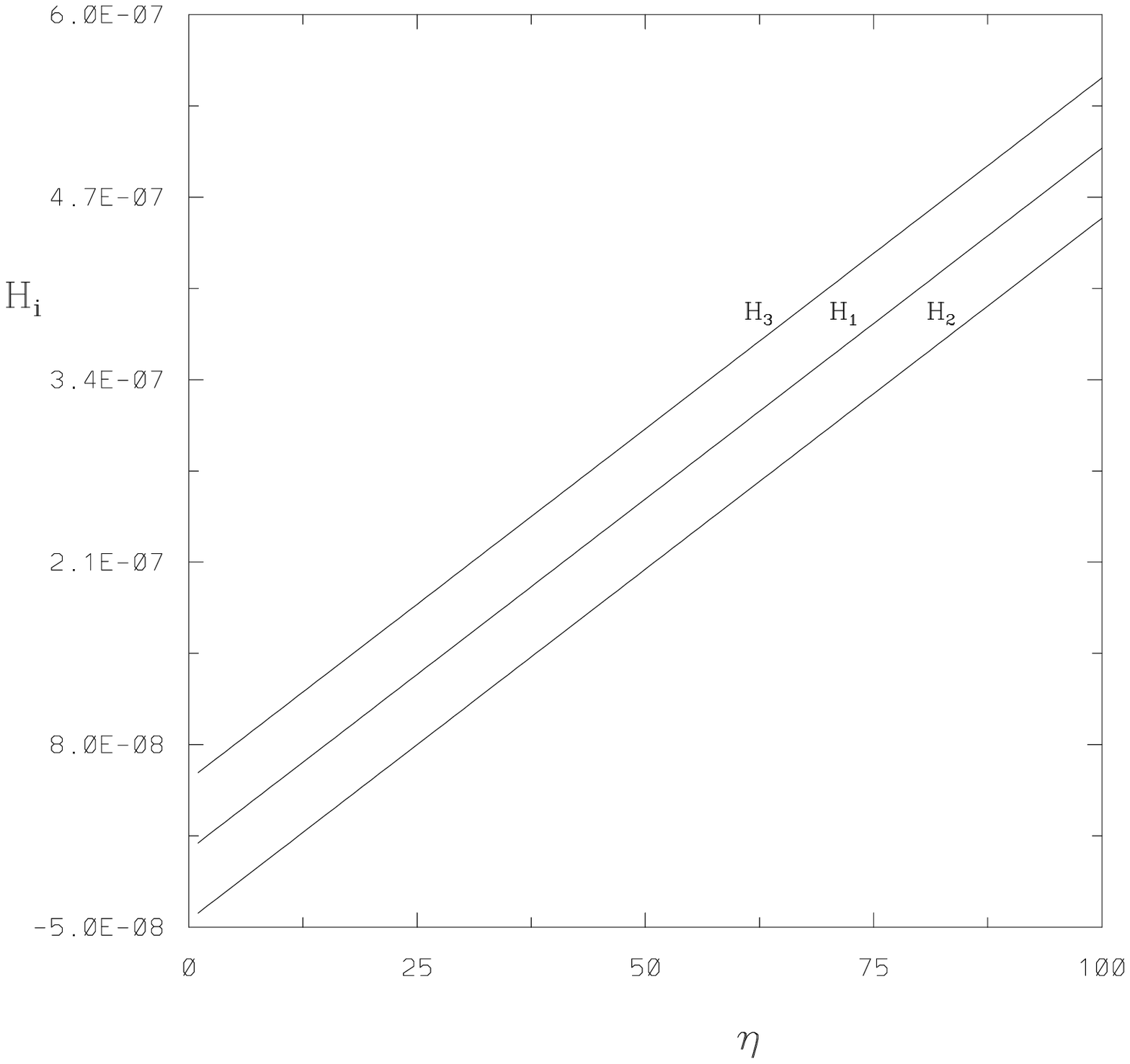}
\caption[The Hubble parameters as a function of time for the Bianchi type 
V (1)]{The Hubble parameters as a function of the time $\eta$.  For these
plots we have taken $\nu=0$, $\omega= 10^{-6} $, and $h_{2}=\eta_{o}=m_{0}=1$,
where the choice of the latter parameters is arbitrary; they are related to
the initial conditions.  In this case, one has that
$H_{1o} > 0$, $H_{3o} > 0$, and  $H_{2o} < 0$.  The latter condition implies
an initial contracting scale factor ($a_{2}$); however, after some evolution
it expands. 
\label{b5i}}
\end{figure}

\begin{figure}
\unitlength 1in
\epsfxsize=5in \epsfbox{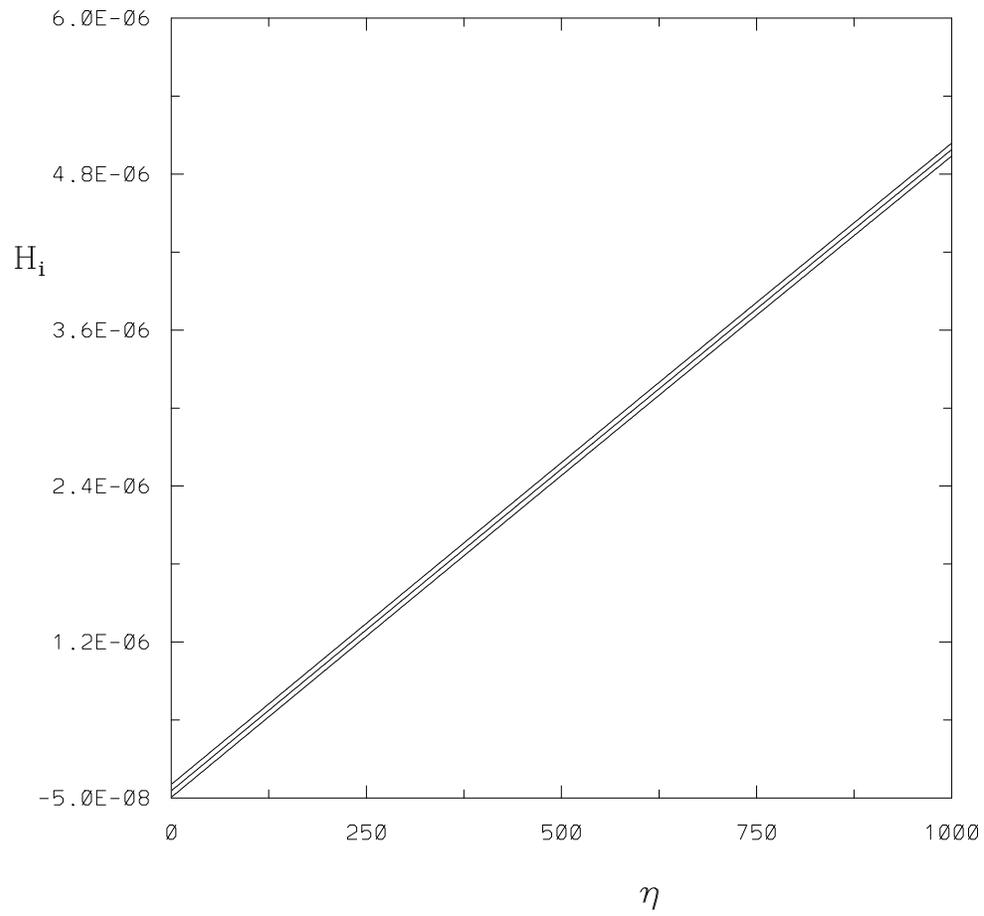}
\caption[The Hubble parameters as a function of time for the Bianchi type V 
(2)]{This figure shows the same as above, but now we plot until the
time $1000$, where one can already observe that the three scale factors
become almost indistinguishable.
\label{b5ii}}
\end{figure}

We plot the Hubble parameters as a function of the time $\eta$ for 
$\omega=10^{-6}$ and $\nu=0$ (corresponding to a dust gas of 
bosons).  Figure \ref{b5i} shows how the anisotropic Hubble parameters evolve 
with the same slop.  This is because the smallness of $\omega$ causes $\psi$ to 
be initially almost a constant, and the $H_i$ solutions are linear, with 
different anisotropic parameters ($h_i$) contributing as different initial 
ordinates, see Eqs. (\ref{hi}). As time elapses the anisotropy becomes 
almost unobservable because of the scale, see figure \ref{b5ii},
making the difference among the three rates of expansion always smaller, that 
is, after some time $\eta_{*}$ the solutions become indistinguishable from the 
open FRW solution, which is given by $H_{1}$, see Ref. \cite{ChCe95}.  Then, as 
mentioned in the previous section, the anisotropy is bounded from above.  In 
Ref. \cite{Ba95} is claimed that the remaining anisotropy is under the limits 
imposed by the COBE satellite on the temperature fluctuations observed in the 
CMBR \cite{Cobe94}.

Note that during the isotropization process, from $\eta_{o}$ to $\eta_{*}$, the
solutions are of the form $H_i \approx D \eta + D_i$, where $D$ and $D_i$ are 
some constants.  Not only the Hubble parameters do not diminish but they 
increase with time! This implies for the scale factors a period of `strong' 
exponential expansion, $a_i \sim e^{D \eta^{2}/2 + D_i \eta}$, causing the 
isotropization of the model.  Note that this solution is very peculiar
and it has its origin in the smallness of $\omega$.  If $\omega$ were not 
that small, a power law solution ($H_{i} \approx 1/\eta$) would be  
valid\footnote{This type of solution is also valid for some set of initial
conditions including any arbitrary value of $\omega$.}.  It is 
curious that we initially aimed to have a model avoiding to begin with 
inflationary initial conditions, and because of the 
smallness of $\omega$ we got an even stronger (than de Sitter) inflationary 
era caused by the Higgs field itself (kinetic + non--minimal coupling),
but not by the potential of the theory.   In this way, during this time the 
Universe begins to loose its {\it anisotropic} hairs because the strong 
exponential expansion dilutes any perturbation present in a distance smaller 
than its event horizon and the no--hair conjecture is then dynamically 
fulfilled.

\begin{figure}
\unitlength 1in
\epsfxsize=5in \epsfbox{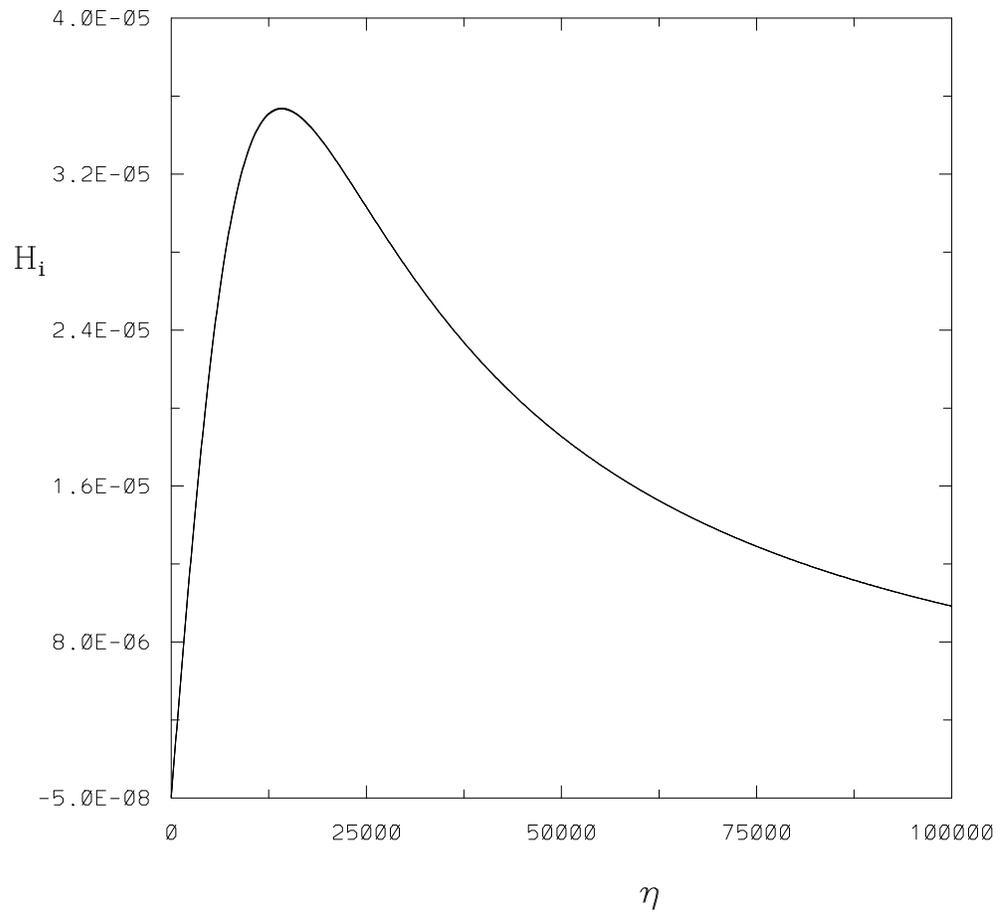}
\caption[The Hubble parameters as a function of time for the Bianchi type V 
model (3)]{The same as above, but now until the time $10^{5}$.  The three
Hubble parameters, here superposed, evolve to an open FRW solution given
by $H_{1}$.
\label{b5iii}}
\end{figure}

\begin{figure}
\unitlength 1in
\epsfxsize=5in \epsfbox{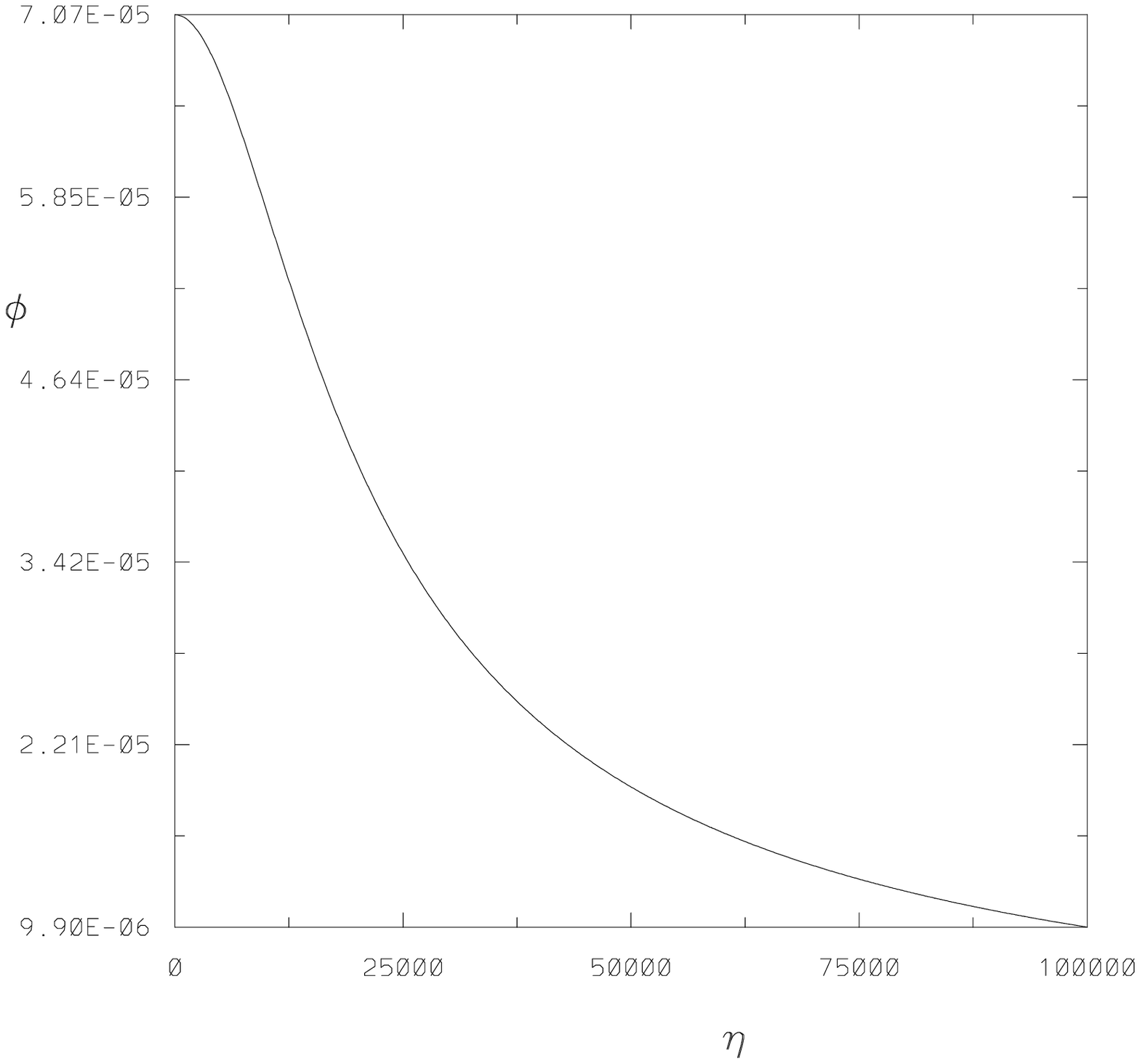}
\caption[The scalar field as a function of time]
{The Higgs field $\phi$ as a function of time, for the same
parameters as above.
\label{phi}}
\end{figure}

In GR the presence of a kinetic term associated to a field (say, $\varphi$) 
added minimally to the Lagrangian, induces the field to evolve 
like $\varphi \sim 1/t^{2}$, see
Ref. \cite{Li90}. Here, we have the same behavior for $\phi$, but this solution
induces an exponential Universe, regardless of the potential. In this case, 
one does not have a slow  rollover dynamics, but the solution itself includes
an exponential expansion with exit (no graceful exit problem) in a natural,
evolutionary way tending to an effective FRW model.  The reason for this is 
that after some elapsed time the variable $\psi$ will be no more effectively 
a  constant and the quadratic term in $\eta$ will dominate and, therefore, 
the solution behaves as shown in figure \ref{b5iii}, that is, with 
$H_{i} \sim 1/\eta$ for $i=1,2,3$.  Figure \ref{phi} shows the evolution 
of the physical Higgs field.

Afterwards, at $\eta=\eta_{1}\ge \eta_{*}$ inflation due to the potential 
takes place.  The initial conditions are such that when inflation is to 
start, the system is already almost isotropic.  Then, inflation (due to the 
potential) begins after the model is to some degree an open FRW model.  The 
dynamics of this stage is studied in Ref. \cite{CeDe95a}.

Figures \ref{b5i}, \ref{b5ii}, \ref{b5iii}, and \ref{phi} describe the dynamics 
dominated by perfect fluid+kinetic terms in early stages, 
precisely when the potential stress energy does not play a significant role in 
the evolution.  But after some elapsed time (at $\eta_{1}$) the potential 
enters into the `game', because the density diminishes rapidly as 
$\rho \sim 1/a^{n}$ with $n=4$ for radiation and $n=3$ for dust like matter 
(boson fields), whereas the potential is only slowly variating.  Thus, from the 
different energy  stresses in Eqs. (\ref{a123}, \ref{h123}, \ref{psi}) the one 
which diminishes slower, as time elapses, is the potential term, in such a way 
that it will eventually be the major stress energy contribution to the 
dynamical equations.  The potential dominance begins\footnote{In this paper 
we assume a chaotic scenario 
for the initial conditions ($\phi_{o}>G^{-1}$, see footnote \ref{f5}), see Refs. 
\cite{CeDe95a,CeDe95b}.} at the $\eta_{1}$ time 
such that $\eta_{1} \ge \eta_{*}$ in order to guarantee some degree of 
isotropization of physical processes present 
before inflation takes place.   Therefore, if any perturbation ({\it hairs}) 
is to be present before that time it must have experienced some degree of 
isotropization, the same as the scale factors.

\begin{figure}[ht]
\unitlength 1in 
\begin{picture}(6,5)(0,1)
\epsfxsize=6in \epsfbox{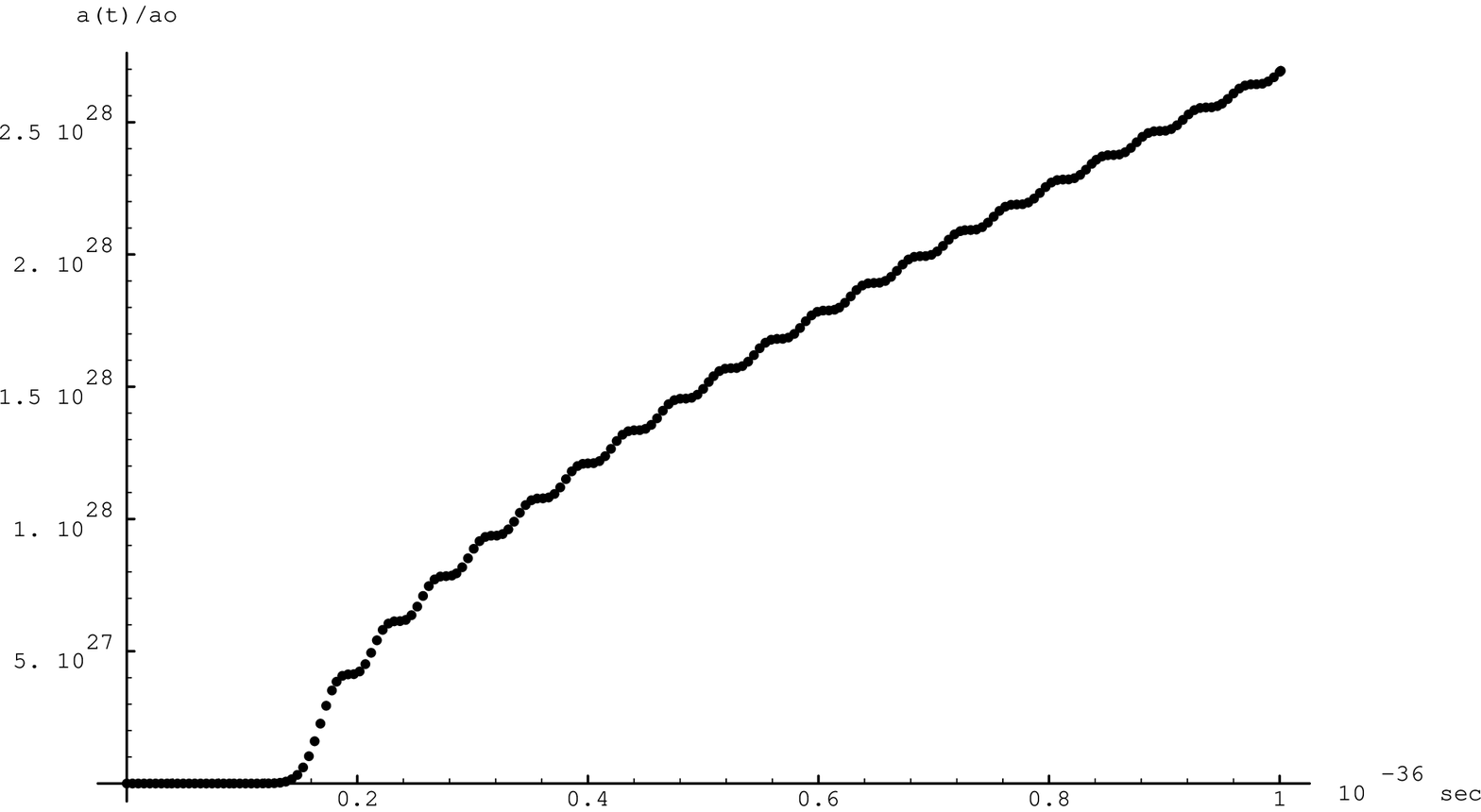}
\end{picture} 
\caption[Inflation and FRW evolution]{The scale factor evolution during and
after inflation until the time $t=10^{2} M^{-1}_{H}$, where $M_{H}$ is the
Higgs mass.  One notes that the inflation time is approximately
$t=2 \times10^{-37} s$, later on, the Universe is  ``dark'' matter dominated
by $\phi$-bosons, perhaps until today, if reheating didn't take away the
coherent Higgs oscillations.  It can be seen the track imprinted by the Higgs
coherent oscillations in the scale factor evolution at that time scale;
afterwards, this influence will be imperceptible.
\label{b5iiii}} 
\end{figure}

After inflation when the Higgs field approaches its symmetry breaking value, 
$\mbox{$ {\rm tr} \Phi^{\dagger} \Phi $}= -6 \mu^{2}/\lambda$, the potential 
diminishes, and begins the high oscillation period  of the Higgs field 
described by Eq. (\ref{hig} or \ref{psi}).  The Higgs oscillations act  
on the scale factors
dynamics with a characteristic frequency given by $M_{H}$ that we have taken
equal to $10^{14}$ GeV.  This high oscillation period induces a FRW model 
($a_i\sim t^{2/3}$) \cite{Tu83}, as it can be observed 
in figure \ref{b5iiii}. Then, inflation, caused by the potential, acts as 
a transient attractor, then, graceful exiting.  This 
result  confirms the general theorems proved in Refs. \cite{Capo95} about the 
no--hair conjecture in scalar tensor theories.  Naturally, one cannot expect 
this solution to be ever trapped in the de Sitter attractor, since the Higgs 
field evolves to its SSB 
minimum, which is the state of lowest energy.  Then, the no--hair conjecture 
fails during the high oscillation period of the Higgs field, as one expects.   

During inflation (due to the potential) perturbations of the Higgs field exit 
the Hubble horizon ($H^{-1}$),
and they will later re--enter to form the seeds of galaxy formation with a 
magnitude\cite{SaBoBa89,Fa90Ma91,CeDe95a}:
\begin{equation}
\label{eqc1}
\frac{\delta\rho}{\rho} \;\vline {\atop{ \atop {}_{\eta_{2}}}} \approx 
\frac{1}{\sqrt{1+\frac{3\alpha}{4 \pi}}} 
H \frac{\delta\phi}{\mbox{$\dot{\phi}$}} 
\;\vline {\atop{ \atop {}_{\eta_{2}}}} =
\sqrt{\frac{1}{6 \pi}}
\frac{M_{H}}{M_{Pl}}  \,\  N(\eta_{2}) \,\ \approx \,\ 
 10 \frac{M_{H}}{M_{Pl}} < 10^{-5} \,\ ,
\end{equation}
where $\eta_{2}$ is the time when the fluctuations of the scalar
field leave $H^{-1}$ during inflation and $N(\eta_{2})$ is the number of 
e-folds of inflation at that time.

The gravitational wave ($GW$) perturbations considered 
normally should also be small \cite{AcZoTu85},
\begin{equation}
\label{eqgw}
h_{GW} \approx \frac{H}{M_{Pl}}  \;\vline {\atop{ \atop {}_{\eta_{2}}}}
\approx   \frac{M_{H}}{M_{Pl}}
\frac{\sqrt{G \phi - 1}}{2} \;\vline {\atop{ \atop {}_{\eta_{2}}}} 
\, \sim  10^{-5} \,\ , 
\end{equation}
which also lies within the experimental limits.

The spectral index of the scalar perturbations, $n_{s}$,
serves as a test for models of the very early universe, independently of the
magnitude of the perturbations.  It can be calculated 
using the slow roll approximation up to second order \cite{KoLi94}.      
For $\omega \ll 1$, however, one can just take the first order to be 
sufficiently accurate \cite{Ka95}:
\begin{equation}
\label{eqnssu5}
n_{s}= 1 - \frac{4}{2 N  + \omega} \approx 1 - \frac{2}{N} \,\ ,
\end{equation}
for $N=65$, it implies $n_{s}\approx 0.97$ in accordance with the recent
COBE DMR results \cite{Cobe94}.

\bigskip

{\bf Bianchi IX}

For the Bianchi IX model the constants 
$A_{{}_{IX}}, B_{{}_{IX}}, C_{{}_{IX}}$ explicit values  are unknown, except
for the value of $A_{{}_{IX}}$ of the isotropic model.  To be physical this 
solution needs to have $\omega<-2$ \cite{ChOb79}.  Therefore, we expect 
to find isotropizing solutions only for $\omega < 0$, but our IG gravity requires 
$\omega>0$ and the scenario is untenable.

\section{Conclusions \label{coniso}}

Within an IG theory we have analyzed Bianchi I, V, and IX models. Only the Bianchi 
type V may isotropize by means of the non--minimal coupling of 
the theory, because the value of $\omega ~(<<1)$ is crucial for
its isotropization.  These results, extracted from the analytic 
solutions in the BD theory \cite{ChCe95}, are here applied to the IG 
theory in an epoch when the potential stress energy is not significant for
the evolution, that is, when both theories are mathematically equivalent.  
If this situation would have happened, initial anisotropies were  
washed out in a Universe with Bianchi V type initial conditions.  On the other 
side, if the potential dominates from the very beginning, inflation 
(because of the potential) occurs directly and it induces the same 
effect, as expected.  

The isotropization mechanism in type V can be inflationary even before the 
potential plays a role or, otherwise, of the $H_{i} \sim 1/\eta$ type.  In the 
former case, from $\eta_{o}$ to $\eta_{1}$, the models experience a strong period 
of exponential expansion, while the Higgs field evolves as $\phi\sim 1/t^{2}$, 
achieving the isotropization of any hairs present (also possibly due to 
some other fields).  After the isotropization mechanism has been concluded, 
because of  Eq. (\ref{hi}), the solution turns away from an exponential behavior 
to become effectively a FRW model, where 
$H_{i}\approx H$ for $i=1,2,3$.  Afterwards, with the Universe ``almost'' isotropic 
the potential begins to dominate the dynamical equations, and therefore 
inflation due to the potential occurs.  Later on,  as the Higgs field 
approaches its ground state value 
($\mbox{$ {\rm tr} \Phi^{\dagger} \Phi $}\rightarrow -6 \mu^{2}/\lambda$), 
inflation ends and an effective FRW dynamics is dominant.  Thus, our 
scenario possesses two inflationary transient attractors, one produced by the 
non-minimal coupling, and the other by the Higgs potential.  The tests of 
inflation for this theory have been proved to be within the 
experimental limits in Refs. \cite{CeDe95a,CeDe95b}.

In the present work, we looked up for the conditions under which isotropization
before inflation occurs in a theory, where the value of $\omega$ is strictly 
determined to be $w=10^{-6}$.   Otherwise, for arbitary  $\omega$--values, the 
mechanism of isotropization can also develop before inflation.  Here
the Bianchi model I tends to the flat FRW solution, the Bianchi model V to the 
open FRW solution, and the Bianchi model IX to the closed FRW solution.

\bigskip

\noindent{\bf Acknowledgments} This work was su\-pported by CONACyT Grant No. 
960021 ``Beca de Repatriaci\'on'' of J.L.C and Grant No. 400200-5-3672E9312 of
P.Ag.C.


\end{document}